# THE NEXT LINEAR COLLIDER KLYSTRON DEVELOPMENT PROGRAM[*]

E. Jongewaard, G. Caryotakis, C. Pearson, R. M. Phillips, D. Sprehn and A. Vlieks
Stanford Linear Accelerator Center, Stanford University, Stanford, CA 94309 USA

*Abstract*

Klystrons capable of 75 MW output power at 11.4 GHz have been under development at SLAC for the last decade. The work has been part of the program to realize all the components necessary for the construction of the Next Linear Collider (NLC). The effort has produced a family of solenoid-focused 50 MW klystrons, which are currently powering a 0.5 GeV test accelerator at SLAC and several test stands, where high power components are evaluated and fundamental research is performed studying rf breakdown and dark current production. Continuing development has resulted in a Periodic Permanent Magnet (PPM) focused 50 MW klystron, tested at SLAC and subsequently contracted for manufacture by industry in England and Japan. A 75 MW version of that PPM klystron was built at SLAC and reached 75 MW, with 2.8 microsecond pulses. Based on this design, a prototype 75 MW klystron, designed for low-cost manufacture, is currently under development at SLAC, and will eventually be procured from industry in modest quantities for advanced NLC tests. Beyond these developments, the design of Multiple Beam Klystrons (MBKs) is under study at SLAC. MBKs offer the possibility of considerably lower modulator costs by producing comparable power to the klystrons now available, at much lower voltages.

## 1 SOLENOID FOCUSED KLYSTRONS

### 1.1 XC Series

SLAC's X-band klystron development began with the XC series of tubes designed to produce 100 MW output power with a relatively high perveance (1.8 µK) electron beam. These klystrons allowed study of several design approaches, including multiple gap output structures and multiple rf output waveguides with separate windows. Unfortunately, the peak power goal proved too high for the existing technology, with rf breakdown in the output cavities and windows, and high voltage breakdown in the electron gun, limiting design operation of the tubes. After a series of six prototypes it was decided to reduce the design goal from 100 MW to 50 MW peak output power for the next series of experimental tubes.

### 1.2 XL Series

This next series of klystrons, designated the XL series, utilized a lower perveance (1.2 µK) than the XC series. Drawing on the results of the XC series, methods to reduce rf fields in critical components were explored. A traveling wave $TE_{01}$ output window was utilized for the first time to reduce rf gradients on the window and eliminate electric fields at the vulnerable window edge. Various multiple gap standing wave and travelling wave output structures were tested, with a four-cell travelling wave output design proving most successful. Vacuum pumping was improved with the addition of an ion pump at the electron gun, effectively reducing the incidence of high voltage arcing in the gun region. The final design variant was designated the XL-4 and has been produced on a limited production basis to power the NLC Test Accelerator and other test facilities at SLAC. This klystron design has operated from 50 MW at 2.4 µs pulse length up to 75 MW at 1.5 µs pulse length. The most recently constructed XL-4 tubes have been further improved by the use of a larger over-moded $TE_{01}$ output window designed for the 75XP-3 PPM klystron. Work is underway to build a modified XL-4 design, designated the XL-X, which will have a larger collector and dual output windows to allow operation at high average power for the purpose of life testing the rf output structure of the klystron.

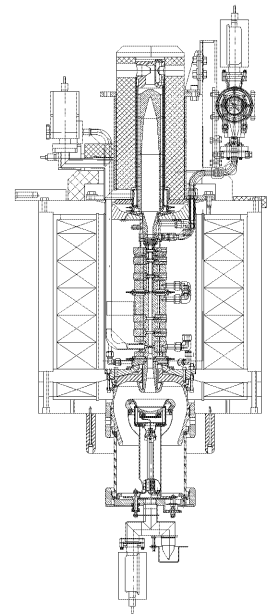

Figure 1: XL-4 Solenoid focused klystron

---

[*] This work supported by the Department of Energy under contract DE-AC03-76SF00515

# 2 PPM FOCUSED KLYSTRONS

## *2.1 Motivation for PPM focusing*

The total number of klystrons required for the NLC is such that any reduction of operating power would lead to great savings in operating costs for the machine. Power savings of some 20 kW per tube could be accomplished by eliminating the solenoid required by the XL type klystrons. This desire to replace the solenoid lead to the idea of using periodic permanent magnet (PPM) focusing, a method used extensively for compact traveling wave tubes, but not on high peak power klystrons. PPM focusing utilizes a periodic focusing field produced by a stack of alternating polarity magnets and field shaping polepieces. If the axial field period is sufficiently small relative to the electron plasma wavelength the beam will be stiffly confined. For this periodic focusing, the rms field value is equivalent to the axial field strength of a solenoidal field.

## *2.2 XL-PPM*

The design for the first PPM klystron began where the XL-4 design left off. The perveance of the beam was dropped to 0.6 μK for greater efficiency and to a make the PPM focusing more effective by increasing the plasma wavelength. To study beam transport without the complication of the rf circuit, a beam diode was designed and constructed. The magnetic design of the diode utilized a gun trim coil and three anode coils that establish the transition field between the cathode region and the PPM magnet stack. These coils allowed adjustment of the gun and transition field during tube operation. The mechanical design of the diode was very similar to that used in PPM focused coupled cavity TWTs. The vacuum envelope was formed by the magnetic circuit itself with alternating iron polepieces brazed to spacers of monel (a non-magnetic alloy of copper and nickel). Like most PPM focused TWTs, the XL-PPM design utilized samarium-cobalt magnets, a rather costly magnet material. Operation of the beam diode proved to be very satisfactory attaining 99.9% beam transmission at 490 kV, 2.8 μs pulse length and a 120 Hz rep-rate, well beyond the design goals for the klystron.

The klystron used a magnetic circuit similar to the diode's with the same matching coil arrangement, but with additional length and a long uni-directional magnet cell to provide focusing for the traveling wave output structure. The circuit construction was the same as the diode's with the addition of rf cavities machined into the monel spacers at the appropriate locations. This construction technique was believed to be beneficial since it provided a lossy surface formed by the alternate iron and monel spacers along the inside of the drift tube to help damp unwanted rf modes that might propagate down the tunnel. The rf design, similar to the successful XL-4 design, also utilized a multi-cell travelling wave output circuit. The reduction of perveance from the XL design necessitated increased drift lengths and the addition of another cell to the output structure. Initial testing of the klystron revealed non-monotonic gain-vs.-drive behavior believed to be caused by multipactor occurring within the drift tube. The klystron was opened and the drift tunnel coated with titanium nitride (TiN) to suppress multipactor. Upon reprocessing and testing the tube was able to operate beyond the design goal and produced peak power over 50 MW for 2.4 μs pulses at a rep rate of 120 Hz; however, operation was limited to two minute bursts by insufficient cooling and excessive x-ray radiation.

## *2.3 75XP-1*

Changes in the design of the NLC lead to the need for a 75 MW rf source, so work began on extending the 50 MW PPM klystron design to this higher power level. Since it was undesirable to increase the beam voltage beyond 500 kV the perveance was increased to 0.75 μK and the beam voltage slightly increased to allow higher beam power for the increased rf requirements. The rf circuit was essentially the same as that of the XL-PPM tube with modifications for the higher perveance and the addition of a cavity for increased gain. The tube construction was somewhat different with stainless steel drift tubes between cavities machined from monel. The magnetic polepieces and spacers were brazed onto these tubes external to the vacuum envelope. It was felt that the stainless drift tube walls would exhibit more rf loss than the drift tunnel of the XL-PPM design and thus better suppress unwanted modes. The magnetic design of the 75XP-1 deviated from that of the XL-PPM in two important ways. First, the anode coils of the XL-PPM were eliminated with the transition field shaped with an all permanent magnet structure. Second, the magnet material was changed to neodymium-iron-boron (Nd-Fe-B), a lower cost, but less temperature stable magnet material. Initial testing of this klystron revealed several instabilities, one of which was identified as a gun oscillation at 1.4 GHz. A loss ring to suppress the oscillation was designed and installed in the electron gun. The klystron was re-tested, the gun oscillation was eliminated, but a 20 GHz oscillation impeded further progress. It was determined that this instability was likely caused by coupling between the output structure and the collector. An rf load was designed and installed in the klystron. The tube was again re-tested and this time was able to produce over 75 MW at 2.8 μs pulse lengths. At this output level the pulse rep rate was limited to 10 Hz due to inadequate cooling of the tube body.

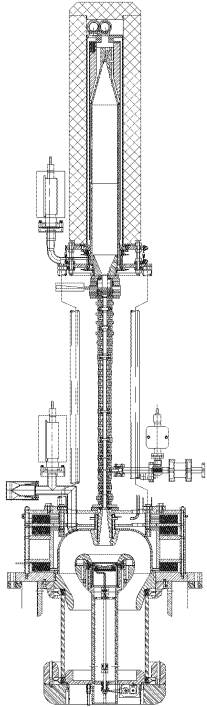

Figure 2: 75XP-3 PPM focused klystron

*2.4 75XP-3*

The SLAC klystron department is currently working on the next device in the series, the 75XP-3. This tube design is intended to address the lessons learned so far, while pushing for full duty operation with a lower cost, simplified design. The new klystron design is very similar to that of the 75XP-1 with refinements to reduce part count, complexity, size and cost. The electron gun is simplified with critical alignment accomplished through precision part machining, eliminating many complex adjustment features found in other SLAC klystrons. Electrode gradients in the electron gun have been designed to be lower than the earlier tubes in an effort to improve reliability. Experience with the fabrication and operation of the prior PPM focused tubes has lead to the conclusion that the ability to test the complete magnetic circuit before tube operation is highly desirable. To accomplish this, a design of a completely separable magnetic structure is underway. This structure will take the form of clam-shell halves containing all the field-forming components that can be built and tested apart from the klystron vacuum envelope. Additionally, the magnet structure can be transferred to another tube and reused as the klystrons reach end of life. Simulations and a test structure have demonstrated that precise fabrication and alignment can control the transverse fields introduced by splitting the magnet structure. Work is ongoing to optimize the thermal design of this clamp-on magnet structure as it serves as the heat sink for any heat dissipation along the klystron drift tube. A beam diode to test the new gun optics and magnetic focusing design is currently under construction and will be tested at the end of this summer. The first klystron of this new design will follow and be tested in the beginning of calendar year 2001.

*2.5 INDUSTRIAL INVOLVEMENT*

The total number of klystrons required for the NLC is such that an industrial source of these tubes is clearly required. However, the SLAC PPM klystrons are unlike any tubes currently produced by industry. To allow commercial tube manufacturers to become familiar with the idiosyncrasies of high peak power PPM klystrons, SLAC has let contracts for the production of XL-PPM 50 MW klystrons to two industrial sources, Marconi in the UK and Toshiba in Japan. These klystrons are currently in production with delivery expected this fall. In addition SLAC has let a contract to CPI in the US for the production of a 75XP-3 rf section and magnet structure that will be mated with electron gun, collector, and output windows supplied by SLAC. By supporting these tube production contracts, SLAC is preparing the microwave tube industry for the large tube orders required for the NLC.

## 3 FUTURE WORK

Work continues at SLAC to further optimize the rf power sources for high-energy collider designs. The successful operation of PPM klystrons was a major milestone in the improvement of overall efficiency of these systems. Other considerations include reducing the cost of the high voltage pulse modulator for the klystron. A reduction of klystron operating voltage will allow consideration of various solid state modulator designs, including direct switching if the operating voltage is low enough. This reduction of operating voltage can be accomplished through the use of multi-beam klystrons where several lower voltage beams are combined in parallel. SLAC is investigating several configurations of multi-beam tube with six to twelve beam devices appearing the most promising.